\documentstyle[12pt,epsf]{article}
 \newcommand{\insertplot}[5]{\begin{figure}
 \hfill\hbox to 0.05in{\vbox to #5in{\vfill
 \inputplot{#1}{#4}{#5}}\hfill}
 \hfill\vspace{-.1in}
 \caption{#2}\label{#3}
 \end{figure}}
 \newcommand{\inputplot}[3]{
 \special{ps: plotfile #1}
}
\newcommand{\vphi}{\varphi}

\begin{document}

\title{
MONOPOLE-ANTIMONOPOLE SOLUTIONS OF EINSTEIN-YANG-MILLS-HIGGS THEORY}
\vspace{1.5truecm}
\author{
{\bf Burkhard Kleihaus}\\
Department of Mathematical Physics, University College, Dublin,\\
Belfield, Dublin 4, Ireland\\
{\bf Jutta Kunz}\\
Fachbereich Physik, Universit\"at Oldenburg, Postfach 2503\\
D-26111 Oldenburg, Germany}

\vspace{1.5truecm}

\date{June 17, 2000}

\maketitle
\vspace{1.0truecm}

\begin{abstract}
We construct static axially symmetric solutions of
SU(2) Einstein-Yang-Mills-Higgs theory in the topologically 
trivial sector, representing 
gravitating monopole--antimonopole pairs,
linked to the Bartnik-McKinnon solutions.
\end{abstract}
\vfill
\noindent {Preprint hep-th/0006148} \hfill\break
\vfill\eject

\section{Introduction}

SU(2) Yang-Mills-Higgs (YMH) theory possesses monopole \cite{mono},
multimonopole \cite{multi,rr,sut},
and monopole-antimonopole pair solutions \cite{taubes,map}.
The magnetic charge of these solutions is proportional
to their topological charge.
While monopole and multimonopole solutions reside in
topologically non-trivial sectors,
the mono\-pole--antimonopole pair solution is topologically trivial.

When gravity is coupled to YMH theory,
a branch of gravitating monopole solutions
emerges smoothly from the monopole solution of flat space \cite{lnw,bfm,lw}.
The coupling constant $\alpha$,
entering the Einstein-Yang-Mills-Higgs (EYMH) equations, 
is proportional to the gravitational constant $G$
and to the square of the Higgs vacuum expectation value $\eta$.
The monopole branch ends at a critical value 
$\alpha_{\rm cr}$,
beyond which gravity becomes too strong 
for regular monopole solutions to persist,
and collapse to charged black holes is expected \cite{lnw,bfm,lw}.
Indeed, when the critical value $\alpha_{\rm cr}$
is reached, the gravitating monopole solutions
develop a degenerate horizon \cite{foot1},
and the exterior space time of the solution
corresponds to the one of an extremal Reissner-Nordstr\o m (RN)
black hole with unit magnetic charge \cite{lnw,bfm,lw,foot2}.

Beside the fundamental gravitating monopole solution, 
EYMH theory possesses radially excited monopole solutions, 
not present in flat space \cite{lnw,bfm,lw}.
These excited solutions also develop a degenerate horizon
at some critical value of the coupling constant,
but they shrink to zero size in the limit $\alpha \rightarrow 0$.
Rescaling of the solutions reveals, that in this limit
the Bartnik-McKinnon (BM) solutions \cite{bm}
of Einstein-Yang-Mills (EYM) theory are recovered.
For the excited solutions the limit $\alpha \rightarrow 0$ 
therefore corresponds to the limit 
of vanishing Higgs expectation value, $\eta \rightarrow 0$.

In this letter we investigate how gravity affects 
the static axially symmetric
mono\-pole--antimonopole pair (MAP) solution of flat space \cite{map},
and we elucidate, that curved space supports a rich spectrum 
of MAP solutions, not present in flat space.

In particular, we show that, with increasing $\alpha$,
a branch of gravitating MAP solutions
emerges smoothly from the flat space MAP solution,
and ends at a critical value $\alpha_{\rm cr}^{(1)}$,
when gravity becomes too strong for regular 
MAP solutions to persist.
But while the branch of monopole solutions can merge into an
extremal RN black hole solution at the critical $\alpha$,
there seems to be no neutral black hole solution
with degenerate horizon available
for the MAP solutions to merge into.
Indeed we find that at $\alpha_{\rm cr}^{(1)}$
a second branch of MAP solutions emerges,
extending back to $\alpha=0$.
Along this upper branch
the MAP solutions shrink to zero size,
in the limit $\alpha \rightarrow 0$,
and approach the BM solution with one node
(after rescaling).

Since the BM solution with one node is related to a branch
of MAP solutions, it immediately suggests itself 
that the excited BM solutions with $k$ nodes are
related to branches of excited MAP solutions.
Indeed, constructing the first excited MAP solution
by starting from the BM solution with two nodes,
we find, that it represents a MAP solution,
possessing two monopole-antimonopole pairs.

\section{\bf Axially symmetric ansatz}

The static axially symmetric MAP solutions of
SU(2) EYMH theory with action
\begin{equation}
S=\int \left ( \frac{R}{16\pi G} 
- \frac{1}{2e} {\rm Tr} (F_{\mu\nu} F^{\mu\nu})
-\frac{1}{4}{\rm Tr}(D_\mu \Phi D^\mu \Phi)
  \right ) \sqrt{-g} d^4x
\   \end{equation}
(with Yang-Mills coupling constant $e$, and vanishing Higgs self-coupling),
are obtained in isotropic coordinates with metric \cite{kk}
\begin{equation}
ds^2=
  - f dt^2 +  \frac{m}{f} \left( d r^2+ r^2d\theta^2 \right)
           +  \frac{l}{f} r^2\sin^2\theta d\vphi^2
\ , \label{metric} \end{equation}
where $f$, $m$ and $l$ are only functions of $r$ and $\theta$.
The MAP ansatz reads for the purely magnetic
gauge field ($A_0=0$) \cite{map}
\begin{equation}
A_\mu dx^\mu = 
\frac{1}{2e}\left\{
\left(\frac{H_1}{r}dr+2(1-H_2)d\theta\right)\tau_\vphi
-2\sin\theta\left(H_3 \tau_r^{(2)}+(1-H_4) 
\tau_\theta^{(2)}\right)d\vphi
\right\}
\label{ansatz} \end{equation}
and for the Higgs field
\begin{equation}
\Phi= \left(\Phi_1 \tau_r^{(2)}+\Phi_2 \tau_\theta^{(2)}\right)
\ , \end{equation}
with $su(2)$ matrices (composed of the standard Pauli matrices $\tau_i$)
\begin{eqnarray}
& \tau_r^{(2)} = \sin 2\theta \tau_\rho +\cos 2\theta\tau_3 \ ,
\ \ \ 
\tau_\theta^{(2)}  = \cos 2\theta \tau_\rho -\sin 2\theta\tau_3 \ , &
\nonumber\\
& \tau_\rho  =  \cos \vphi \tau_1+ \sin \vphi\tau_2\ ,
\ \ \ 
\tau_\vphi   =  -\sin \vphi \tau_1+ \cos \vphi\tau_2\ .
\end{eqnarray}
The four gauge field functions $H_i$ and the two Higgs field functions
$\Phi_i$ depend only on $r$ and $\theta$.
We fix the residual gauge degree of freedom \cite{rr,kk,map} 
by choosing the gauge condition 
$r\partial_r H_1-2\partial_\theta H_2 =0$ \cite{map}.

To obtain regular asymptotically flat solutions
with finite energy density
we impose at the origin ($r=0$) the boundary conditions
\begin{eqnarray}
& H_1=H_3=H_2-1=H_4-1=0\ , &
\nonumber \\
& \sin 2\theta \Phi_1 + \cos 2\theta \Phi_2 =0\ , \ \ \ 
\partial_r
\left(\cos 2\theta \Phi_1 - \sin 2\theta \Phi_2\right) = 0
\ , &
\nonumber\\
& \partial_r f = \partial_r m = \partial_r l=0 \ . &
\nonumber
\end{eqnarray}
On the $z$-axis the functions 
$H_1, H_3, \Phi_2$ and the derivatives
$\partial_\theta H_2,\partial_\theta H_4,\partial_\theta \Phi_1 ,
\partial_\theta f, \partial_\theta m,\partial_\theta l$ have to vanish,
while on the $\rho$-axis the functions
$H_1, 1-H_4, \Phi_2$ and the derivatives
$\partial_\theta H_2,\partial_\theta H_3,\partial_\theta \Phi_1 ,
\partial_\theta f, \partial_\theta m,\partial_\theta l$ have to vanish.
For solutions with vanishing net magnetic charge the gauge potential
approaches a pure gauge at infinity. 
The corresponding boundary conditions 
for the fundamental MAP solution are given by \cite{map}
\begin{equation}
H_1=H_2=0 \ ,
H_3=\sin \theta \ , 1-H_4=\cos\theta \ , 
 \Phi_1 = \eta \ , \Phi_2=0 \ , f=m=l=1 \ .
\end{equation}

Introducing the dimensionless coordinate 
$x=r\eta e$ and the Higgs field $\phi = \Phi/\eta$,
the equations depend only on the coupling constant $\alpha$,
$\alpha^2 = 4\pi G\eta^2$.
The mass $M$ of the MAP
solutions can be obtained directly from
the total energy-momentum ``tensor'' $\tau^{\mu\nu}$
of matter and gravitation,
$M=\int \tau^{00} d^3r$ \cite{wein},
or equivalently from
$ M = - \int \left( 2 T_0^{\ 0} - T_\mu^{\ \mu} \right)
   \sqrt{-g} dr d\theta d\phi $,
yielding the dimensionless mass $\mu = \frac{4\pi\eta}{e} M$.

\section{\bf Solutions}

Subject to the above boundary conditions,
we solve the equations numerically \cite{xgmap}.
In the limit $\alpha \rightarrow 0$, 
the lower branch of gravitating MAP solutions 
emerges smoothly from the flat space solution \cite{map}.
The modulus of the Higgs field of these MAP solutions
possesses two zeros, $\pm z_0$, on the $z$-axis,
corresponding to the location of the monopole and antimonopole, 
respectively.

\begin{figure}\centering\epsfysize=8cm
\mbox{\epsffile{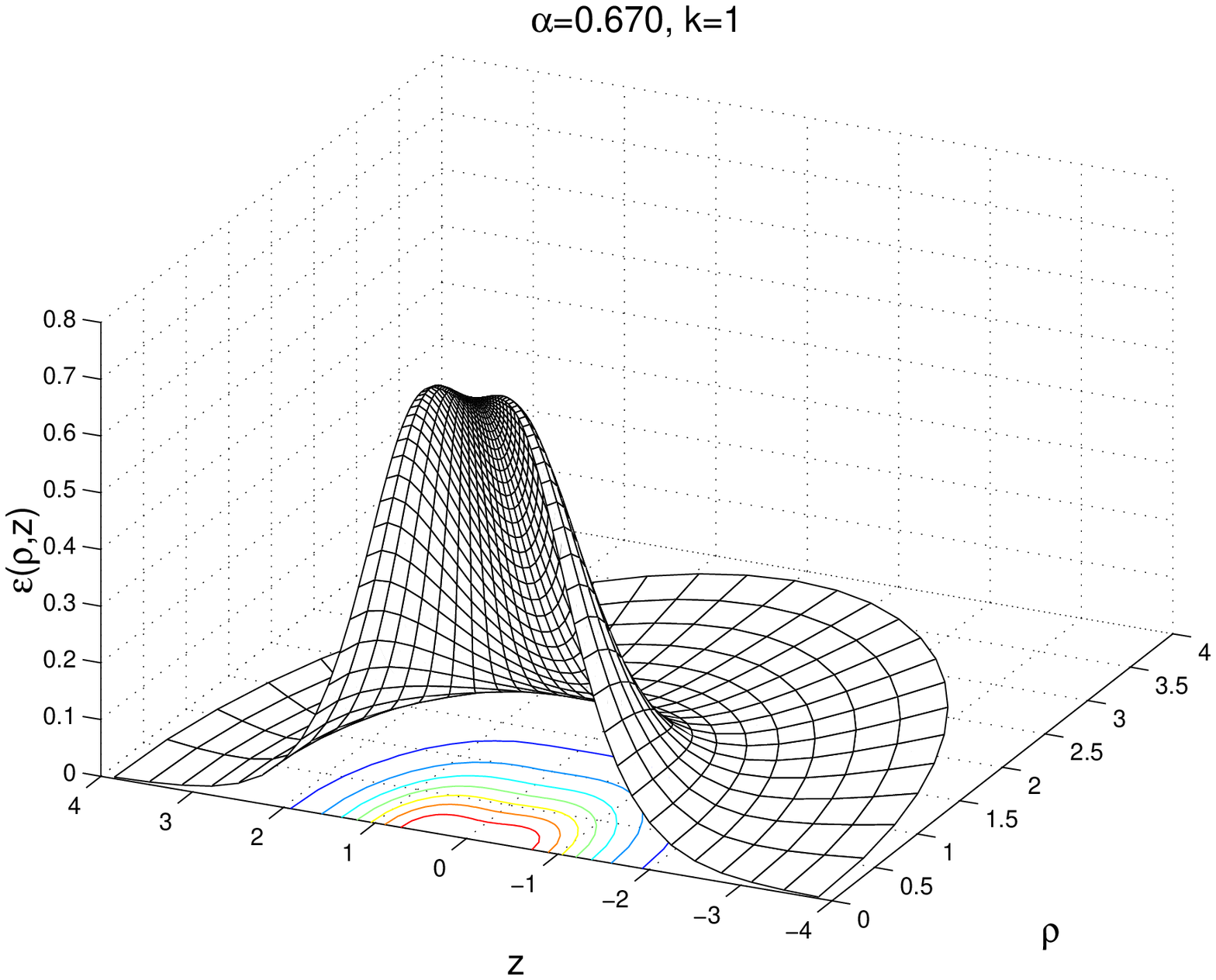}}
\caption{\label{eps}
The energy density $\varepsilon(\rho,z)$ is shown for the 
fundamental MAP solution at $\alpha_{\rm cr}^{(1)}=0.67.$}
\end{figure}

With increasing $\alpha$ 
the monopole and antimonopole move closer to the origin,
and the mass $\mu$ of the solutions decreases.
The lower branch of MAP solutions ends
at the critical value $\alpha_{\rm cr}^{(1)}=0.670$.
In Fig.~\ref{eps} we show the energy density 
$\varepsilon = -T_0^0 = -L_M$
of the MAP solution at $\alpha_{\rm cr}^{(1)}$.
It possesses maxima on the positive and negative $z$-axis
close to the locations of 
the monopole and antimonopole 
and a saddle point at the origin.

\begin{figure}\centering\epsfysize=8cm
\mbox{\epsffile{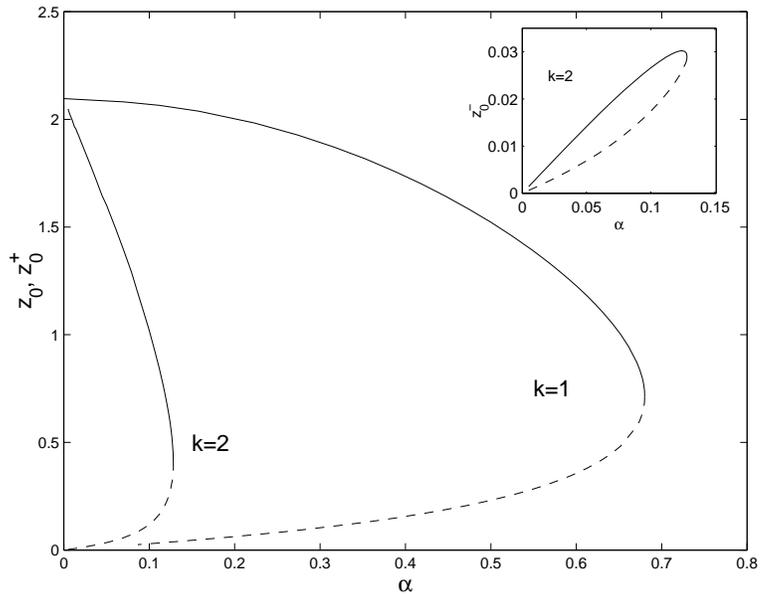}}
\caption{\label{z02}
For the fundamental ($k=1$) and the first excited ($k=2$) MAP solution
the locations of the monopole, $z_0$ resp.~$z_0^+$, are shown as 
functions of $\alpha$. 
In the inlet the location of the antimonopole, $z_0^-$, of the 
first excited MAP solution is shown.
The solid and dashed lines correspond to  
the lower and upper (mass) branches, respectively.}
\end{figure}

Forming a second branch,
the MAP solutions evolve smoothly backwards
from $\alpha_{\rm cr}^{(1)}$ to $\alpha=0$.
In the limit $\alpha \rightarrow 0$
the mass $\mu$ diverges on this upper branch,
and the locations of the monopole and antimonopole
approach the origin, $\pm z_0 \rightarrow 0$,
as seen in Fig.~\ref{z02}.
At the same time the MAP solution shrinks to zero. 

\begin{figure}\centering\epsfysize=8cm
\mbox{\epsffile{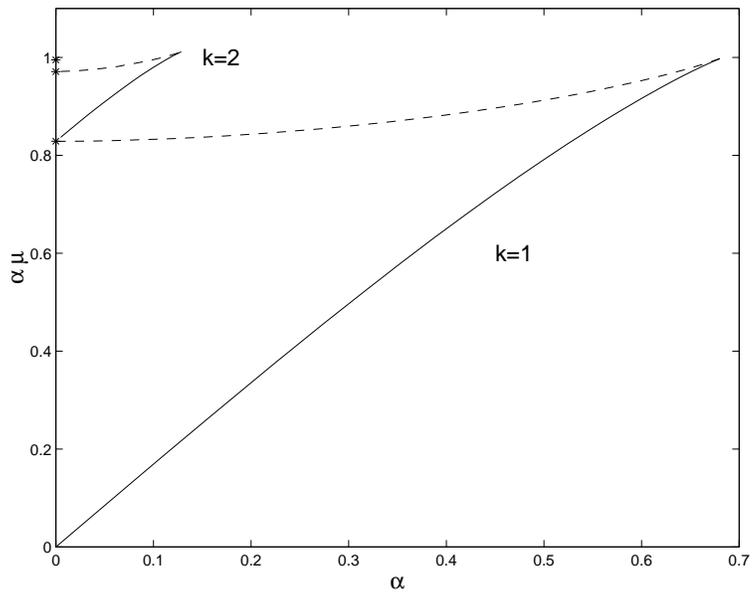}}
\caption{\label{E}
The scaled mass $\hat \mu=\alpha \mu$ is shown 
as a function of $\alpha$ for the fundamental ($k=1$) and 
the first excited ($k=2$) MAP solution.
The solid and dashed lines correspond to
the lower and upper (mass) branches, respectively.
The stars indicate the masses of the $k=1,2,3$ (from bottom to top)
BM solutions.}
\end{figure}

Rescaling the coordinate $x= \hat{x}\alpha$ 
and the Higgs field $\phi = \hat{\phi}/\alpha$
reveals that the axially symmetric MAP solutions 
approach the spherically symmetric $k=1$ BM solution
on the upper branch as $\alpha \rightarrow 0$. 
Consequently, also the scaled mass $\hat{\mu}= \alpha \mu$ 
of the MAP solutions tends 
to the mass of the $k=1$ BM solution, as seen in Fig.~\ref{E}.
On the upper branch the limit $\alpha \rightarrow 0$ 
thus corresponds to the limit $\eta \rightarrow 0$
(with fixed $G$).
We note that the ansatz (\ref{ansatz})
for the gauge potential includes
the spherically symmetric BM ansatz,
\begin{equation}
H_1 = 0 \ , \ \ 
1-H_2 =  \frac{1}{2}(1-w) \ , \ \ 
H_3= \frac{1}{2}\sin\theta (1-w)) \ , \ \
1-H_4 =  \frac{1}{2} \cos\theta (1-w)  \ ,
\end{equation} 
where $w$ denotes the gauge field function of the BM solution.

Anticipating the existence of excited MAP solutions,
linked to the BM solutions with $k$ nodes on their upper branches,
we construct the first excited MAP solution,
starting from the $k=2$ BM solution.
Since the boundary conditions of the $k=2$ BM solution 
differ from those of the $k=1$ BM solution at infinity,
the boundary conditions of the first excited MAP solution 
at infinity must be modified accordingly,
\begin{equation}
 H_1=H_3=0\ , \ \ \ H_2=H_4=1\ , \ \ \ 
 \phi_1 = \pm \cos 2\theta \ , \ \ \ \phi_2 = \mp \sin 2\theta \ , 
\ \ \  f=m=l=1 \ .
\end{equation}

The upper branch of the first excited MAP solutions
ends at the critical value  $\alpha_{\rm cr}^{(2)} = 0.128 $,
from where the lower branch of the excited MAP solutions
evolves smoothly backwards to $\alpha=0$.
As seen in  Fig.\ref{E}, 
in the limit $\alpha \rightarrow 0$ the scaled mass $\hat{\mu}$
approaches the mass of the $k=2$ BM solution on the upper branch,
and the mass of the $k=1$ BM solution on the lower branch.

The modulus of the Higgs field of the first excited MAP solution
possesses four zeros, $\pm z_0^+$ and $\pm z_0^-$, 
located on the $z$-axis,
representing two monopole-antimonopole pairs.
The locations of the monopole and antimonopole on the positive $z$-axis, 
$z_0^+$ resp.~$z_0^-$, are  shown in Fig.~\ref{z02} 
as functions of $\alpha$, together with 
the node $z_0$ of the fundamental MAP solution.
As $\alpha \rightarrow 0$,  
$z_0^-$ tends to zero on both branches;
in contrast, $z_0^+$ tends to zero only on the upper branch.
On the lower branch $z_0^+$ tends to $z_0$,
the location of the monopole of the fundamental MAP solution.

Inspecting the limit $\alpha \rightarrow 0$ 
for the first excited MAP solution on the lower branch reveals,
that in terms of the radial coordinate $x=r\eta e$, 
the solution differs from the fundamental MAP solution 
on its lower branch only near the origin, 
where the excited MAP solution develops a discontinuity.
In terms of the coordinate $\hat{x}=x/\alpha$, on the other hand, 
the first excited MAP solution approaches the $k=1$ BM solution 
for all values of $\hat{x}$, except at infinity.
Hence, the first excited MAP solution does not possess 
a counterpart in flat space.

\section{Conclusions}

Having constructed the fundamental and the first excited MAP solutions,
we expect, that EYMH theory possesses a whole sequence
of MAP solutions, labeled by the number of monopole-antimonopole pairs $k$.
Each MAP solution forms two branches, 
merging and ending at $\alpha_{\rm cr}^{(k)}$.
In the limit $\alpha \rightarrow 0$,
the upper branch of the $k$th MAP solution always reaches the
Bartnik-McKinnon solution with $k$ nodes,
while the lower branch of the $k$th MAP solution always reaches the
Bartnik-McKinnon solution with $k-1$ nodes,
except for $k=1$, where the flat space MAP solution is reached
in the limit $\alpha \rightarrow 0$.
We conjecture, 
that the critical values $\alpha_{\rm cr}^{(k)}$ decrease with $k$,
such that, as a function of $\alpha$,
the scaled mass $\hat \mu$ assumes
a characteristic ``Christmas tree'' shape.
Thus instead of the single MAP solution present in flat space,
in curved space a whole tower of MAP solutions appears.
An analogous pattern is encountered for gravitating Skyrmions, which are
likewise linked to the BM solutions \cite{ESk}.
We expect the graviating MAP solutions to be unstable 
like the flat space MAP solution \cite{taubes}.

For the gravitating monopole solutions a regular event horizon
can be imposed \cite{lnw,bfm,lw}, yielding 
magnetically charged black hole solutions with hair.
Likewise for the MAP solutions of EYMH theory
a regular event horizon can be imposed,
yielding static axially symmetric
and neutral black hole solutions with hair \cite{inprep}.
Within the framework of distorted isolated horizons the masses
of these black hole solutions may possibly be simply related
to the masses of the corresponding regular solutions 
\cite{ashtekar}.

It is interesting, that the spherically symmetric BM solutions of EYM
theory appear in the limit $\alpha \rightarrow 0$
of the axially symmetric MAP solutions.
But EYM theory also possesses static axially symmetric regular solutions,
which are not spherically symmetric \cite{kk}.
Could these solutions also appear in the $\alpha \rightarrow 0$ limit
of more general \cite{brikunz} gravitating MAP solutions?
We conjecture, that EYMH theory allows for the existence
of MAP solutions, consisting of pairs of static axially symmetric
multimonopoles, where each multimonopole has winding number $n$
\cite{multi,rr}.
It is then conceivable that such multimonopole-antimultimonopole solutions
will form an analogous set of solutions as 
the ones encountered above, but
with their upper branches reaching axially symmetric EYM solutions
with winding number $n$ in the $\alpha \rightarrow 0$ limit.

But also flat space should contain further interesting solutions,
for instance an antimonopole-monopole-antimonopole system,
with the poles located symmetrically with respect to the origin
on the $z$-axis.

\clearpage
\newpage

\begin{thebibliography}{000}
\bibitem{foot2}
 For large Higgs self-coupling 
 the exterior solution is not of RN-type \cite{lw}.
\end{thebibliography}
\end{document}